\documentclass[journal=jpccck,manuscript=article]{achemso}

\usepackage[version=3]{mhchem} 

\usepackage{amsmath}
\usepackage{epsf}
\usepackage{epsfig}
\usepackage[T1]{fontenc}
\usepackage[latin9]{inputenc}
\usepackage{array}
\usepackage{multirow}

\author{C. Ataca}
\altaffiliation{UNAM-Institute of Materials Science and
Nanotechnology, Bilkent University, Ankara 06800, Turkey}
\affiliation{Department of Physics, Bilkent University, Ankara
06800, Turkey}
\author{H. \c{S}ahin}
\affiliation{UNAM-Institute of Materials Science and
Nanotechnology, Bilkent University, Ankara 06800, Turkey}
\author{E. Akt\"{u}rk}
\affiliation{UNAM-Institute of Materials Science and
Nanotechnology, Bilkent University, Ankara 06800, Turkey}
\author{S. Ciraci}\email{ciraci@fen.bilkent.edu.tr}
\affiliation{UNAM-Institute of Materials Science and
Nanotechnology, Bilkent University, Ankara 06800, Turkey}
\affiliation{Department of Physics, Bilkent University, Ankara
06800, Turkey}

\title[MoS2]
  {Mechanical and Electronic Properties of MoS$_2$ Nanoribbons and Their Defects}

\abbreviations{IR,NMR,UV}

\begin{tocentry}
\includegraphics{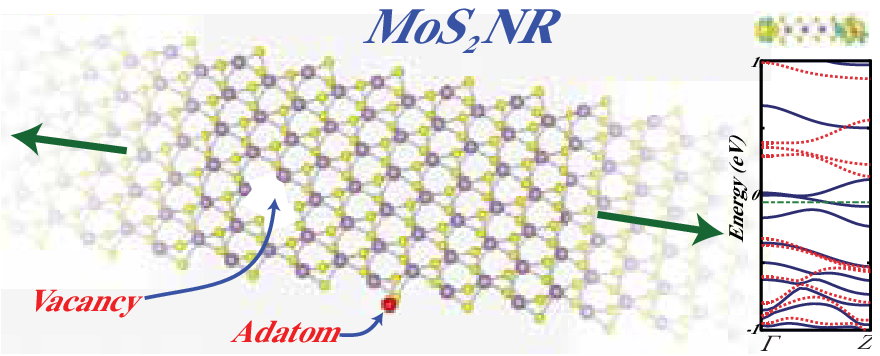}

\end{tocentry}

\date{\today}

\begin{document}

\begin{abstract}

We present our study on atomic, electronic, magnetic and phonon properties of one dimensional honeycomb structure of molybdenum disulfide (MoS$_2$) using first-principles plane wave method. Calculated phonon frequencies of bare armchair nanoribbon reveal the fourth acoustic branch and indicate the stability. Force constant and in-plane stiffness calculated in the harmonic elastic deformation range signify that the MoS$_2$ nanoribbons are stiff quasi one dimensional structures, but not as strong as graphene and BN nanoribbons. Bare MoS$_2$ armchair nanoribbons are nonmagnetic, direct band gap semiconductors. Bare zigzag MoS$_2$ nanoribbons become half-metallic as a result of the (2x1) reconstruction of edge atoms and are semiconductor for minority spins, but metallic for the majority spins. Their magnetic moments and spin-polarizations at the Fermi level are reduced as a result of the passivation of edge atoms by hydrogen. The functionalization of MoS$_2$ nanoribbons by adatom adsorption and vacancy defect creation are also studied. The nonmagnetic armchair nanoribbons attain net magnetic moment depending on where the foreign atoms are adsorbed and what kind of vacancy defect is created. The magnetization of zigzag nanoribbons due to the edge states is suppressed in the presence of vacancy defects.
\end{abstract}

\section{Introduction}

Unique honeycomb orbital symmetry underlies the unusual properties of two dimensional (2D) single hexagonal structures, such as graphene,\cite{novo,zhang,berger} silicene and group III-IV binary compounds.\cite{ansiklopedik} Moreover, quasi 1D nanoribbons and flakes of these 2D layers have added interesting electronic and magnetic properties, which are expected to give rise to important future applications in nanotechnology.\cite{graphene_applications1,graphene_applications2,graphene_applications3,graphene_applications4,sahinJAP} Recently, 2D suspended single layer molybdenum disulfide, MoS$_2$ sheets with honeycomb structure have been produced.\cite{Joenson,novo2} Single layer MoS$_2$ nanocrystals of $\sim$ 30 \AA~ width were also synthesized on the Au(111) surface and its direct real space STM have been reported.\cite{Helveg} Unlike graphite and hexagonal BN, the layers of MoS$_2$ are made of hexagons with Mo and S$_2$ atoms situated at alternating corners. Apparently, 3D graphitic bulk structure called 2H-MoS$_2$, 2D single layer called 1H-MoS$_2$, quasi 1D nanotubes\cite{Seifert} and nanoribbons of MoS$_2$ share the honeycomb structure and are expected to display interesting dimensionality effects.

Properties of MoS$_2$ nanocrystals are explored in diverse fields, such as nanotribology,\cite{rapoport,lee} hydrogen production,\cite{Hinnemann,Jaramillo} hydrodesulfurization catalyst used for removing sulfur compounds from oil,\cite{HDS1,HDS2,HDS3,HDS4,HDS5,HDS6,HDS7} solar cells,\cite{Kline} photocatalysis.\cite{Wilcoson} Triangular MoS$_2$ nanocrystals were obtained as a function of size by using atom-resolved scanning tunneling microscopy.\cite{Lauritsen} Photoluminescence emerging from 1H-MoS$_2$ was observed.\cite{galli} Superlow friction between surfaces coated with 1H-MoS$_2$ has been measured much recently.\cite{lee} Using electrochemical methods micro and nanoribbons have been synthesized from crystalline 2H-MoS$_2$.\cite{qLi} Various properties of 2H-MoS$_2$ (see Ref.[ \cite{Wieting, Kasowski, Mattheiss, Wakabayashi, Coehoorn, Batsonov, Kobayashi, Frey, Boker, Schroder, Fuhr, Ivanovskaya, Huang, Moses, Mendez, Sun}]), 1H-MoS$_2$ (see Ref.[ \cite{Bertrand, Sandoval, Seifert, Li, Lebegue, Mendez, Bollinger}]) and its nanoribbons (see Ref.[\cite{Li, Mendez}]) have been an active subject of theoretical studies.

In this paper, we present our systematic theoretical investigation of optimized atomic structure and phonon spectrum, mechanical, electronic, magnetic properties of armchair (A-MoS$_2$NR) and zigzag (Z-MoS$_2$NR) nanoribbons. Our study reveals interesting results, which are important for further study and applications of these nanoribbons. These are: (i) We demonstrated the stability of MoS$_2$ nanoribbons through first-principles calculations of phonon frequencies. Specifically we deduced the branch of twisting modes. (ii) We calculated force constants and in-plane stiffness of armchair and zigzag nanoribbons showing that they are stiff materials. (iii) We examined the effects of the reconstruction of edge atoms and their passivation by hydrogen on the electronic and magnetic properties of nanoribbons. The energy is optimized through a $(2 \times 1)$ reconstruction of edge atoms of zigzag nanoribbons, that, in turn, renders half-metallicity. (iv) The properties of MoS$_2$ nanoribbons can be dramatically modified by foreign atom adsorption and vacancy defects. Since recent works\cite{lee, mak, Jaramillo} show that single layer MoS$_2$ flakes as large as 200 $\mu$m$^2$ can now be produced and also can be characterized, present results are crucial for further research on MoS$_2$ nanoribbons.

\section{Method}

Our results are based on first-principles plane wave calculations within density functional theory (DFT) using projector augmented wave (PAW) potentials.\cite{paw} The exchange correlation potential is approximated by generalized gradient approximation (GGA) using PW91\cite{pw91} functional both for spin-polarized and spin-unpolarized cases. All structures are treated using the periodic boundary conditions. Kinetic energy cutoff, Brillouin zone (BZ) sampling are determined after extensive convergence analysis. A large spacing of $\sim$ 10 \AA~ between the S planes of two MoS$_2$ layer are taken to prevent interactions. A plane-wave basis set with kinetic energy cutoff of 600 eV is used. In the self-consistent field potential and total energy calculations BZ is sampled by special \textbf{k}-points by using Monkhorst-Pack scheme.\cite{Monk} For nanoribbons, BZ is sampled by 1x1x9 \textbf{k}-points. All atomic positions and lattice constants are optimized by using the conjugate gradient method, where the total energy and atomic forces are minimized. The convergence for energy is chosen as 10$^{-5}$ eV between two consecutive steps, and the maximum Hellmann-Feynman forces acting on each atom is less than 0.05 eV/\AA~upon ionic relaxation. The pressure in the unit cell is kept below 1 kBar. The phonon dispersion curves are calculated along symmetry directions of BZ within density functional theory using Small Displacement Method (SDM).\cite{alfe} Numerical calculations have been performed by using VASP.\cite{vasp1,vasp2} Bader analysis is used for calculating the charge on adatoms.\cite{Bader}

\section{Properties of Two Dimensional MoS$_2$}

For the sake of comparison we first present a brief discussion of the properties of 2D 1H-MoS$_2$ calculated with the same parameters used for quasi 1D nanoribbons. Single layer MoS$_2$ structure consists of monatomic Mo plane having a 2D hexagonal lattice, which is sandwiched between two monatomic S planes having the same 2D hexagonal lattice. Mo and S$_2$ occupy alternating corners of hexagons of honeycomb structure. The contour plots of calculated charge density and difference charge density isosurfaces clarify the charge distribution in layers of MoS$_2$ structure. Electronic charge transferred from Mo to S atoms gives rise to an excess charge of 0.205 electrons around each S atom.\cite{Mulliken, Pauli} This situation implies that 1H-MoS${_2}$ can be viewed a positively charged Mo planes between two negatively charged planes of S atoms and this is the main reason why flakes of MoS$_2$ structure are good lubricant. The cohesive energy of 1H-MoS$_2$ is calculated as 15.55 eV using GGA+PAW. The structure is optimized to yield hexagonal lattice constant, $a=$3.20~\AA, and internal structure parameters, such as the bond distance between Mo and S atoms $d_{Mo-S}$=2.42~\AA, the distance between two S atoms at each corner $d_{S-S}$=3.13~\AA, and the angle between Mo-S bonds $\Theta_{S-Mo-S}$=80.69$^{o}$. The ground state of monolayer 1H-MoS$_2$ is nonmagnetic semiconductor having direct band gap of $E_{g}$=1.58 eV. The upper part of the valence and the lower part of the conduction bands are dominated from bonding and antibonding Mo-$4d$ and S-$3p$ orbitals.

\section{1D MoS$_2$ Nanoribbons}

Two dimensional 1H-MoS$_2$ can maintain its physical properties, when its size is large. However a small flake or a ribbon can display rather different electronic and magnetic properties. In particular, edge atoms may influence the physical properties. The passivation of edge atoms by hydrogen atoms also result in significant changes in the properties of the nanoribbons. In this respect, one expects that the armchair (A-MoS$_2$NR) or zigzag (Z-MoS$_2$NR) nanoribbons of 1H-MoS$_2$ can display even more interesting electronic and magnetic properties.

We consider bare, as well as hydrogen saturated armchair and zigzag nanoribbons. These nanoribbons are specified by their width $w$ in \AA~ or $n$ number of Mo-S$_2$ basis in the unit cell. We take armchair nanoribbon with $n$=12 and zigzag nanoribbons with $n$=6 as prototypes. The distance between Mo and S atom, $d_{Mo-S}$ varies depending on the position in the ribbon. For example for $n$=12, while at the center of the armchair nanoribbon, $d_{Mo-S}$=2.42 \AA, and $d_{S-S}$=3.13 \AA, at the edge of the armchair nanoribbons, they change to $d_{Mo-S}$=2.56 \AA~and $d_{ S-S}$=3.27 \AA. The lattice parameters at the center of ribbons attain the same values as 1H-MoS$_{2}$. The average binding energy of hydrogen atoms passivating Mo and S atoms at the edges of the nanoribbon is $E_b$=3.64 eV. The lengths of Mo-H and S-H bonds are 1.70 \AA~ and 1.36 \AA, respectively. The distance between S atoms at the edge is calculated as 3.27 \AA~ upon hydrogen passivation.

\subsection{Phonon calculations, stability and elastic properties}

\begin{figure}
\centering
\includegraphics[width=8cm]{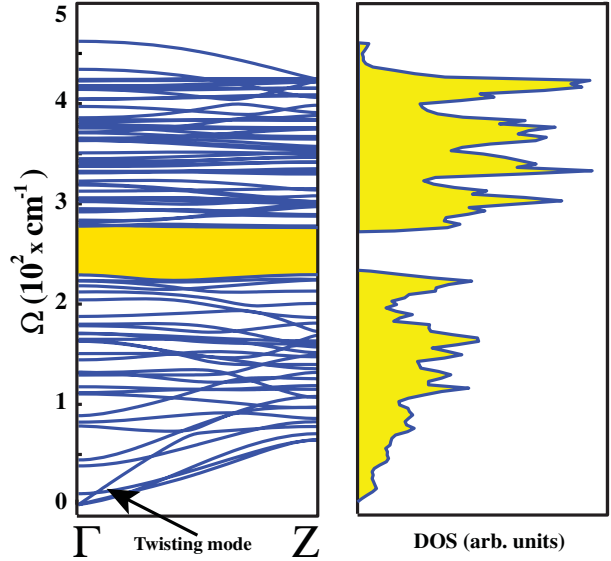}
\caption{(Color online) Calculated phonon frequencies, $\Omega$(\textbf{k}) of the bare armchair MoS$_2$ nanoribbon with $w$= 17.75 \AA ~or $n$=12 (there are $36$ atoms in the primitive cell) are presented along symmetry directions of the Brillouin zone using Small Displacement Method (SDM), and corresponding densities of states (DOS).}
\label{fig:phonon}
\end{figure}

While the structure optimization through energy minimization yields an indication whether a given nanoribbon is stable, a rigorous test for the stability can be achieved through phonon calculations. If the calculated frequencies of specific phonon modes are imaginary, the structure becomes unstable for the corresponding \textbf{k}-wave vectors in the BZ. Here we present an example for the stability test of nanoribbons, whereby we calculate the phonon frequencies of the bare armchair nanoribbon with $n$=12. The calculated phonon branches and corresponding density of states (DOS) are presented in ~\ref{fig:phonon}. The out of plane (ZA) branch with parabolic dispersion and fourth acoustic branch (or twisting mode\cite{seymurtwist}) characteristics of nanoribbons are obtained. Earlier, the branch of twisting mode was revealed in MoS$_2$ nanotubes.\cite{twist1} Similar twisting modes are also calculated for ZnO nanoribbons.\cite{twist2} The overall shape of DOS of nanoribbons are similar to that of 2D MoS$_2$ sheets\cite{Lee2}, except that the gap between optical and acoustical branches is reduced due to edge phonon states. For the same reason additional peaks occur for flat phonon branches of edge modes in band continua. All modes having positive frequency indicate that the bare armchair nanoribbon of MoS$_2$ with $n$=12 is stable. It is also expected that other bare nanoribbons having $n >$ 12 are stable.

Having demonstrated the stability of nanoribbons, we next investigate their mechanical properties by calculating the elastic properties. Currently, the behavior of honeycomb structures under tension has been a subject of current interest.\cite{straineffecta,straineffectb,straineffectc,kesme,carbonchain,topsakal_PRB} While 1H-MoS$_{2}$ have honeycomb structure, its atomic configuration and bonding of atoms are dramatically different from graphene. Therefore, the response of A- and Z-MoS$_2$NRs to the strain is expected to be different from graphene. The elastic properties of the quasi 1D MoS$_2$ nanoribbons are examined through the variation of the total energy with applied strain. Generally, nanoribbons change their electronic and magnetic properties under uniaxial tension in the elastic deformation range. Here we present the response of A-MoS$_2$NR and Z-MoS$_2$NR to the strain in elastic range.

Nanomechanics of both bare A-MoS$_2$NR with $n$=12 and Z-MoS$_2$NR with $n$=6 is explored by calculating the mechanical properties as a response to the strain along the axis of the ribbon. To allow more variational freedom and reconstruction, segments of these NRs are treated within supercell geometry using periodic boundary conditions and spin-polarized calculations are carried out. Each supercell, both having total of 108 atoms, contains three unit cells for armchair and six unit cells for zigzag nanoribbons, respectively. The stretching of the ribbon is achieved by increasing the equilibrium lattice constant $c_{0}$ by $\Delta c$, to attain the axial strain $\epsilon=\Delta c/c_{0}$. We optimized the atomic structure at each increment of the strain, $\Delta\epsilon=$0.01 and calculated the total energy under strain $E_{T}(\epsilon)$. Then the strain energy can be given by, $E_{S}=E_{T}(\epsilon)-E_{T}(\epsilon=0)$; namely, the total energy at a given strain $\epsilon$ minus the total energy at zero strain. The tension force, $F_{T}=-\partial E_{S}(\epsilon)/\partial c$ and the force constant $\kappa=\partial^{2}E_{S}/\partial c^{2}$ are obtained from the strain energy. Owing to ambiguities in defining the Young's modulus of honeycomb structures, one can use in-plane stiffness $C=(1/A_{0})\cdot(\partial^{2}E_{S}/\partial\epsilon^{2})$ in terms of the equilibrium area of the supercell, $A_{0}$.\cite{yakobson1,reddy} The in-plane stiffness can be deduced from $\kappa$ by defining an effective width for the ribbon.

For both A- and Z-MoS$_2$NR the hexagonal symmetry is disturbed, but overall honeycomb like structure is maintained in the elastic range. However, stretched ribbons can return to its original geometry when the tension is released. In the harmonic range the force constant is calculated to be $\kappa$= 116.39 N/m and 92.38 N/m for A-MoS$_2$ having $n=$12 and Z-MoS$_2$NR having $n=$6, respectively. Similarly, the calculated in-plane stiffness for the same ribbons are $C$=108.47 N/m and 103.71 N/m, respectively. The difference between the values of armchair and zigzag nanoribbon occurs due to different bond and edge directions. As the width of the nanoribbon goes to infinity these two values are expected to converge to a single value. The calculated values are smaller than the values of $C$=292 and 239 N/m calculated for graphene and BN honeycomb armchair nanoribbons.\cite{topsakal_PRB} Nevertheless, both calculated $\kappa$ and $C$ values indicate the strength of 1H-MoS$_2$. It should be noted that $\kappa$ is approximately proportional to $n$, but $C$ is independent of $n$ for large $n$. Small deviations arise from the edge effects.

For applied strains in the plastic deformation range the atomic structure of the ribbon undergoes irreversible structural changes, whereby uniform honeycomb structure is destroyed. At the first yielding point the strain energy drops suddenly, where the ribbons undergo an irreversible structural transformation. Beyond the yielding point the ribbons are recovered and started to deform elastically until next yielding. Thus, variation of the total energy and atomic structure with stretching of nanoribbons exhibit sequential elastic and yielding stages.

\subsection{Electronic and magnetic properties}

Similar to the single layer 1H-MoS$_2$, its armchair nanoribbons (A-MoS$_2$NR) are also semiconductors. The bare A-MoS$_2$NR is a nonmagnetic, direct band gap semiconductor. Upon hydrogen termination of the edges, the band gap increases. Also the direct band gap shows variation with $n$, like the family behavior of graphene nanoribbons. However, unlike armchair graphene nanoribbons\cite{cohen}, the band gaps of A-MoS$_2$NR's do not vary significantly with its width $w$ or $n$. For narrow armchair nanoribbons with $n<7$ the calculated value of the band gap is larger than that of wide nanoribbons due to quantum confinement effect. The variation of $E_g$ with $n$ is in agreement with that calculated by Li \emph{et.al}.\cite{Li}

\begin{figure}
\centering
\includegraphics[width=8cm]{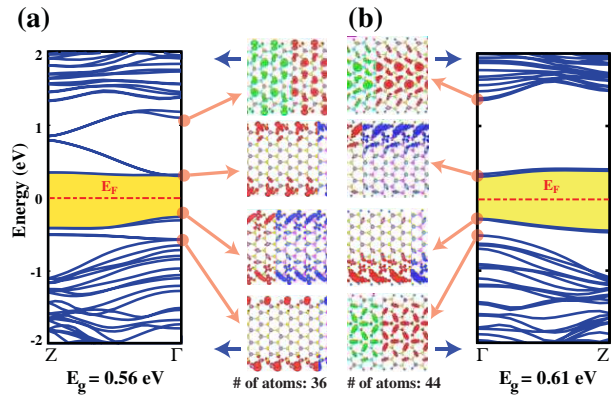}
\caption{(Color online) (a) Energy band structure of bare A-MoS$_2$NR
having $n$=12 and the width $w$=17.75 \AA. The band gap is shaded and the zero
of energy is set at the Fermi level. At the right-hand side, charge density isosurfaces of specific states at the conduction and valence band edges are shown. (b) Same as (a), but the edge atoms are saturated by H atoms as described in the text. Large (purple), medium (yellow) and small (blue) balls are Mo, S, and H atoms,
respectively. Short and dark arrows indicate the direction of the axes of nanoribbons. Total number of atoms in the unit cells are indicated.}

\label{fig:armchair}
\end{figure}

The electronic band structure and charge density of specific states are examined in detail for a bare A-MoS$_2$NR of $n$=12 in ~\ref{fig:armchair}(a). The edge states, which are driven from Mo-$4d$ and S-$3p$ orbitals and have their charge localized at the edges of the nanoribbon form flat bands located in the large band gap of 1H-MoS$_2$. Because of these edge states, the band gaps of bare armchair nanoribbons are smaller than that of 1H-MoS$_2$. Upon termination of each Mo atom at the edge by two hydrogen and each S atom by a single hydrogen atom, the part of edge states are discarded and thus the band gap slightly increases. As seen in ~\ref{fig:armchair} (b), the remaining edge states continue to determine the band gap of the ribbon. Even if the character of these bands change, their charges continue to be located near the edge of the ribbon. Nevertheless, the band gaps of hydrogen saturated armchair nanoribbons remain to be smaller than that of 2D 1H-MoS$_2$.

\begin{figure}
\centering
\includegraphics[width=8cm]{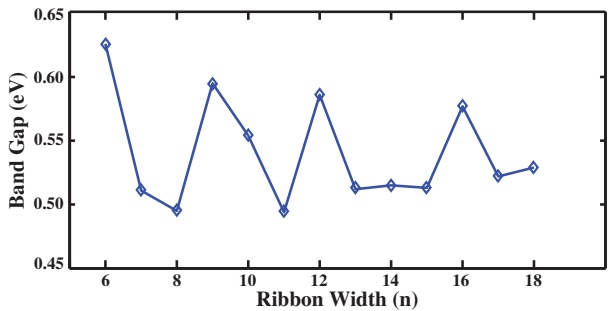}
\caption{(Color online) Variation of direct band gap of hydrogen saturated armchair nanoribbons, A-MoS$_2$NR with $n$. Each Mo atom at the edge is passivated by two hydrogen atoms and each S atom is passivated by a single H atom. }

\label{fig:harmchair}
\end{figure}

Furthermore we investigated the variation of band gap of hydrogen saturated armchair nanoribbons as a function of $n$. As shown in~\ref{fig:harmchair} for $n\leq$ 7 the values of band gap are larger due to quantum confinement effect, but for $n \geq$ 7 they tend to oscillate showing a family like behavior. These oscillations follow those found for bare armchair nanoribbon.\cite{Li} All calculated A-MoS$_2$NR are found to be direct band gap semiconductors.

\begin{figure}
\centering
\includegraphics[width=8cm]{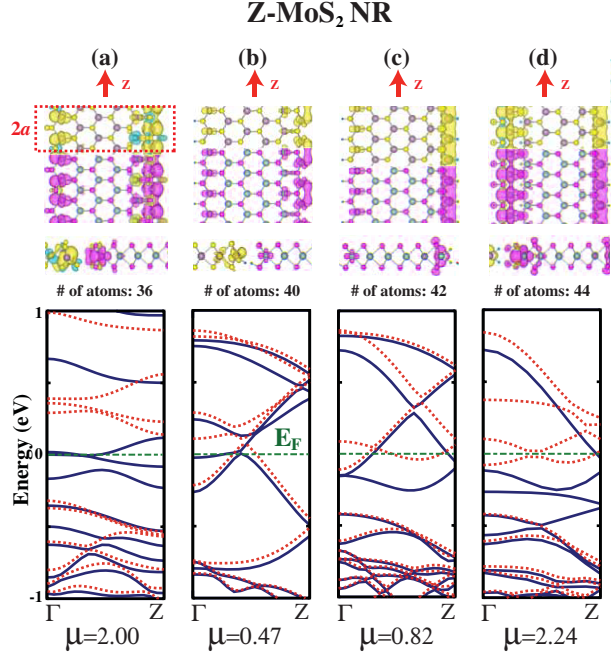}
\caption{(Color online) Atomic and energy band structure of bare and hydrogen saturated zigzag nanoribbon Z-MoS$_2$NR having $n$=6 Mo-S$_2$ basis in the primitive unit cell. The top and side views of the atomic structure together with the difference of spin-up and spin-down charges, $\Delta \rho = \rho^{\uparrow}-\rho^{\downarrow}$, are shown by yellow/light and turquoise/dark isosurfaces, respectively. The isosurface value is taken to be 10$^{-3}$ electrons per \AA$^{3}$. The (2x1) unit cell with the lattice constant $2a$ is delineated. Large (purple), medium (yellow) and small (blue) balls are Mo, S, and H atoms, respectively. The zero of energy is set at the Fermi level shown by dash-dotted green/dark lines. Energy bands with solid (blue) and dashed (red) lines show spin-up and spin-down states, respectively. (a) The bare Z-MoS$_2$NR having $\mu$=2 $\mu_B$ per cell displays half-metallic properties. (b) Spin-polarized ground state of Z-MoS$_2$NR with Mo atoms at one edge and bottom S atoms at the other are passivated by single hydrogen. (c) Similar to (b), but Mo atoms are passivated by two hydrogen atoms. (d) Similar to (c), but top S atoms at the other edge are also passivated by single hydrogen atoms. The net magnetic moment of each case is indicated below the corresponding band panels. Bands are calculated using double cells. Small arrows along $z$-axis indicate the direction of the nanoribbon. The total number of atoms in supercell calculations are indicated for each case.}

\label{fig:zigzag}
\end{figure}

In contrast to A-MoS$_2$NR, the bare zigzag nanoribbons Z-MoS$_2$NR are spin-polarized metals. Here we consider Z-MoS$_2$NR with $n$=6 as a prototype. In \ref{fig:zigzag}, it is shown that the edge atoms of this nanoribbon undergo a $(2 \times 1)$ reconstruction by lowering its total energy by 0.75 eV. Interestingly, as a result of reconstruction, the bare Z-MoS$_2$NR is a half-metal with integer magnetic moment per primitive cell, namely $\mu$=2 $\mu_B$. Thus the nanoribbon is metallic for majority (spin-up) bands, but a semiconductor for minority (spin-down) bands with an indirect band gap of $\sim 0.50$ eV. We check that half-metallic state of bare Z-MoS$_2$NR's is maintained for $n$=5, and $n$=8. Half-metals are interesting spintronic materials and were revealed first in 3D crystals.\cite{groot} Lately, various nanostructures, such as Si nanowires\cite{engin2a} and atomic chains of carbon-transition metals compounds\cite{sdag} have found to display half-metallic properties. The half-metallic property is destroyed upon the saturation of the edge atoms by hydrogen. The magnetic moment of the ribbon and the density of spin states at the Fermi level depend on how Mo and S atoms at the edges of the ribbon are passivated by hydrogen. One distinguishes three different hydrogen passivations, each leads to different magnetic moments as indicated in \ref{fig:zigzag}. As the number of passivating hydrogen atoms increases the number of bands crossing the Fermi level decreases. However the spin-polarization is relatively higher, if each S atoms at one edge are passivated by single hydrogen atom and each Mo atom at the other edge is passivated by double hydrogen. Interestingly, the latter nanoribbon in~\ref{fig:zigzag}(d) is metallic for one spin direction and semimetal for the opposite spin direction. Different spin polarizations found for different spin directions can make potential nanostructure for applications in spintronics.

Earlier Li \emph{et al.}\cite{Li} examined electronic and magnetic properties of armchair and zigzag MoS$_2$ nanoribbons using VASP\cite{vasp1,vasp2} within DFT. They found that armchair nanoribbons are nonmagnetic semiconductors and their direct energy band gap vary with $n$ and becomes 0.56 eV as $n\rightarrow \infty$. They did not consider hydrogen passivation of edge atoms. They also noted that the value of net magnetic moment can change, but the ferromagnetic state of zigzag nanoribbons are maintained even after H passivation of edge atoms. Mendez \emph{et al.}\cite{Mendez} investigated armchair nanoribbons and concluded that these nanoribbons are metallic and have a net magnetic moment, but they change to semiconductor after hydrogen passivation of edge atoms. Their calculations show that in the case of bare armchair nanoribbons, the magnetic state is energetically more favorable by 14 meV and for H-saturated zigzag nanoribbons the antiferromagnetic state is favorable relative to the ferromagnetic state by 15 meV. These results disagree the present results, as well as with those of Li \emph{et al.}\cite{Li}

Normally, the bare and unreconstructed zigzag nanoribbons have sizable electric dipole moment along the direction from the edge having only negatively charged S atoms to the other edge having only positively charged Mo atoms. The dipole moment is calculated to be 55.4 e\AA ~per cell of Z-MoS$_2$NR having $n=6$, but it reduces to 0.07 ~\AA~ upon reconstruction of the edges. Present results show that the edge reconstruction ought to be treated properly to reveal the half-metallic state and to estimate the correct dipole moment.

\subsection{Impurities and Defects in MoS$_2$ Nanoribbons}

Interesting properties of MoS$_2$ nanoribbons revealed above can be modified through adatom adsorption (or doping) and vacancy defect creation. Earlier, Huang and Cho\cite{Huang} investigated the adsorption of CO on a pure 1H-MoS$_2$ surface by using DFT. Similarly, aromatic and conjugated compounds on MoS$_2$ are also studied.\cite{Moses} Similar to graphene\cite{graphene_applications3,esquinazi,Iijima,yazyev,guinea,brey2,sahingraphane,canBN,ethemSiGe,canc} and its nanoribbons,\cite{brey1,delik,sahinPRB,canc} not only adatoms, but also vacancy defects created can led to crucial effects.

\begin{center}
\begin{figure}
\includegraphics[width=8cm]{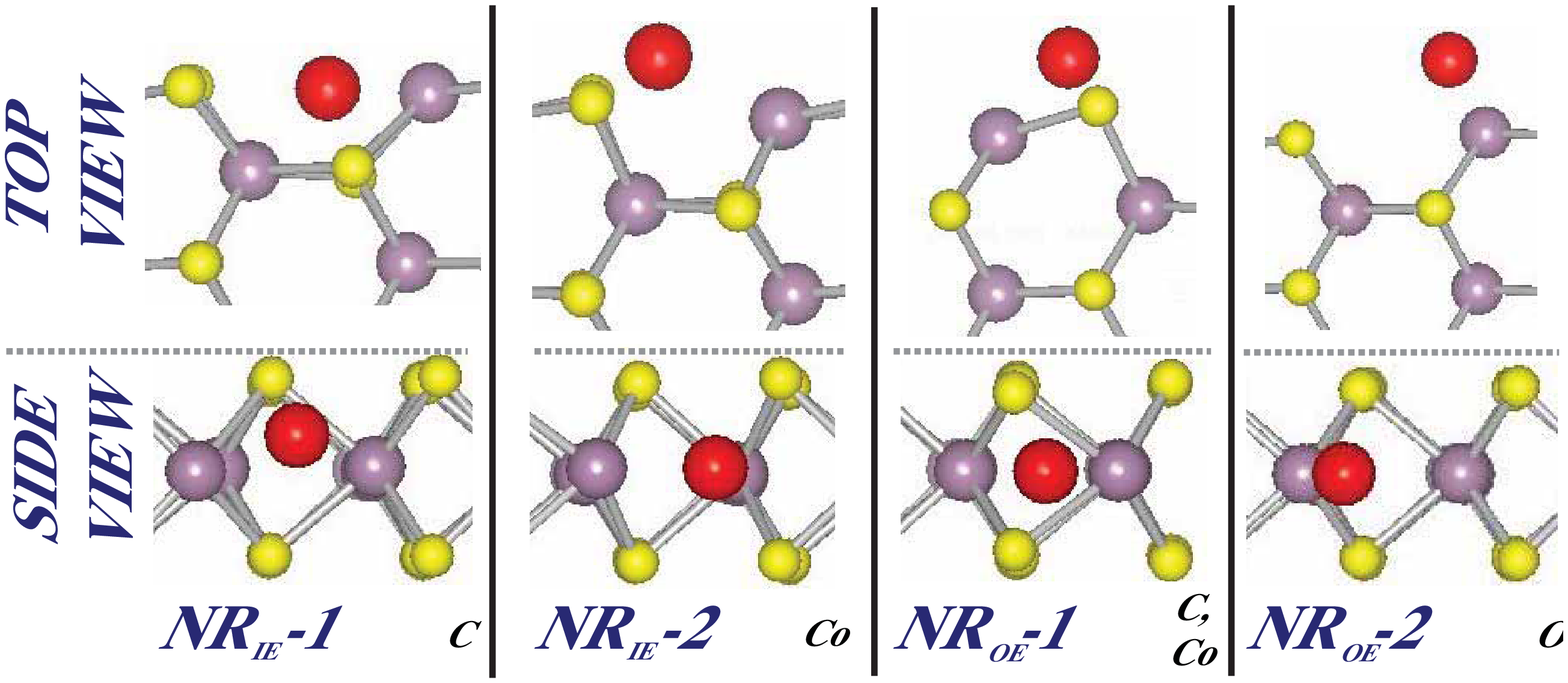}
\caption{(Color online) Top and side views for the schematic representation of possible adsorption geometries of adatoms obtained after the structure optimization. Adatoms, Mo and S are represented by red (large-dark), purple (medium-gray) and yellow (small-light) balls, respectively. Side view clarifies the height of adatoms from Mo and S atomic planes. In each possible adsorption geometry, the entry on the lower-left part indicates where the adatom is initially placed. All sites show geometries associated with the adsorption to a bare armchair (n=12) nanoribbon (NR). The calculations are carried out in the supercell geometry where a single adatom is adsorbed at every three unit cells. The total number of atoms in the supercell is $109$. Possible adsorption geometries in NR$_{IE}$ (adatom is initially placed at the inner edge of bare armchair NR) and NR$_{OE}$ (adatom is initially placed at the outer edge of bare armchair NR). Adatoms indicated at lower right part of every possible adsorption geometry correspond to those, which are relaxed to this particular geometry upon structure optimization.}
\label{fig4}
\end{figure}
\par\end{center}

\begin{table*}
\caption{Calculated values of adatoms adsorbed to the bare armchair MoS$_2$ nanoribbon having $n$=12 MoS$_{2}$ units in the primitive unit cell. The supercell in calculations consist of three primitive cells. There are two different adsorption sites as (described in \ref{fig4}) for each adatoms. The positions only with a positive binding energy is indicated. $h_{Mo}$, the height of the adatom from Mo layer; $h_{S}$, the height of the adatom from the nearest S-layer; $d_{Mo}$, the adatom-nearest Mo distance; $d_{S}$, the adatom-nearest S distance; $E_b$, adatom binding energy; $\mu_{T}$, magnetic moment per supercell in Bohr magneton $\mu_{B}$; $\rho^{*}$, excess charge on the adatom (where negative sign indicates excess electrons); $\Phi$, photoelectric threshold (work function); P, dipole moment calculated in the $x$, $y$ and $z$ direction, respectively. Nanoribbon is in the $(x,y)$-plane and along the $x$-direction. $E_{i}$, energies of localized states induced by adatoms. Localized states are measured from the top of the valence bands in electron volt. The occupied ones are indicated by bold numerals and their spin alignments are denoted by either $\uparrow$ or $\downarrow$. States without indicated spin alignment are nonmagnetic.}
\label{adatom}
\tiny
\begin{tabular}{ccccccccccccc}
\hline \hline \tabularnewline
  Atom & Site & $h_{Mo}$ &
$h_{S}$ & $d_{Mo}$ & $d_{S}$ &
 $E_{b}$  & $\mu_{T}$  & $\rho^{*}$ & $\Phi$ & P & E$_i$
 \tabularnewline

 & & (\AA) & (\AA) & (\AA) & (\AA) & (eV) & ($\mu_{B}$) & (e) & (eV) & (e
$\times$  \AA) & $\uparrow$ : Spin-up, $\downarrow$ : Spin-down States
\tabularnewline \hline \hline

 \multirow{2}{*}{C}

& NR$_{IE}$-1 & 0.63 & 0.95 & 1.95 & 1.81 & 5.69 & NM & -0.67 & 5.60
& (-22.45, 0.63, -0.15) & \textbf{-15.05}, \textbf{-9.05}
\tabularnewline

& NR$_{OE}$-1  & -0.01 & 1.55 & 2.00 & 1.79 & 6.35 & NM & -0.56 &
5.43 & (-26.58, 0.52, -0.02) & \textbf{-15.63}, \textbf{-8.48},
\tabularnewline

\hline

 \multirow{1}{*}{O}

& NR$_{OE}$-2 & 0.00 & 1.56 & 1.72 & 3.56 & 6.67 & NM & -0.73 & 5.64
& (0.84, 1.52, 0.00) & \textbf{-5.82}, \textbf{-5.81}, \textbf{-5.63}, \textbf{-1.16}, \textbf{-0.90}
\tabularnewline

\hline

 \multirow{2}{*}{Co}

& NR$_{IE}$-2 & 0.11 & 1.45 & 2.40 & 2.15 & 4.81 & 1.00 & 0.22 &
5.42 & (3.15, 0.04, -0.09) & \textbf{-1.12}$\uparrow$, \textbf{-0.40}$\uparrow$, \textbf{-0.36}$\downarrow$, \textbf{-0.31}$\uparrow$, 0.38$\downarrow$, 0.38$\downarrow$, 1.23$\uparrow$
\tabularnewline

& NR$_{OE}$-1 & -0.02 & 1.55 & 2.32 & 2.17 & 4.44 & 0.85 & 0.26 &
5.19 & (-13.83, -1.45, -0.03) & \textbf{-0.95}$\uparrow$, \textbf{0.16}$\uparrow$, \textbf{0.29}$\downarrow$, 0.56$\downarrow$, 1.17$\downarrow$, 1.23$\uparrow$,
\tabularnewline

\hline \hline

\end{tabular}
\label{tab:NR}
\end{table*}

 Here we consider again our prototype armchair nanoribbon and investigate the adsorption of C, O and Co. C is widely investigated in other honeycomb structures; the adsorption O is expected to result in important changes due to its high electronegativity. On the other hand, Co being a transition metal atom is expected to attribute magnetic properties. In the supercell geometry, a single adatom is adsorbed at every three unit cells, which leads to the adatom-adatom distance of $\sim 16.60$ \AA. We found that the edges of the nanoribbon are active sites for adsorption and are energetically more favorable relative to the center of nanoribbon. As described in ~\ref{fig4}, adatoms adsorbed at the inner (NR$_{IE}$) and outer parts of the edges (NR$_{OE}$) of the armchair nanoribbon result in a reconstruction on the edges and form strong bonds with nanoribbon. In~\ref{tab:NR}, we present all relevant data obtained from our calculations of adatoms adsorbed to A-MoS$_2$NR. The height of the adatom from the Mo- or S-planes are calculated relative to the average heights of Mo- and S- atoms in the corresponding planes. The binding energy, $E_{b}$ is calculated as $E_{b} = E_{ad} + E_{A-MoS_2} - E_{ad+A-MoS_2}$. Here, $E_{ad}$ is the ground state energy of free adatom calculated in the same supercell with the same parameters; $E_{A-MoS_2}$ is the total energy of nanoribbon and $E_{ad+A-MoS_2}$ is the energy of adatom+A-MoS$_2$NR complex. The charge at the adatom $\rho_{B}$, is calculated using Bader analysis.\cite{Bader} The excess charge of the adatom is obtained by subtracting the charge at the adatom, $\rho_{B}$ from the valence charge of the adatom Z$_{A}$, namely $\rho^{*}=Z_{A}-\rho_{B}$. Accordingly $\rho^{*}$ < 0 implies excess electron at the adatom site. Here the adatom+A-MoS$_2$NR complex attains net magnetic moment after the adsorption of transition metal atom, Co. Adsorption of C and O do not cause any spin polarization in all adsorption geometries. Adsorbed O having the highest electronegativity among the adsorbates treated here has highest excess charge; C is also negatively charged in both adsorption geometries. Co adatom having electronegativity smaller than those of both constituent atoms of the nanoribbon is positively charged. The depletion and annihilation of charge from the adatom result in a small dipole moment in the $y$-direction, which is normal to the ribbon. Since adatom-adatom interaction is hindered due to large supercell dimensions, the localized states form flat bands in the supercell geometry. For C and O localized states deep in the valance band are generally due to their low energy $2s$-states. For Co adatom most of the localized states originate from $3d$-orbitals.

 We note that previously, He\cite{He} \emph{et al.} found that the lowest energy adsorption position of C adatom is at the top of Mo atom in monolayer MoS$_2$ and oxygen adatom is adsorbed to the top site above the S atom. However, adsorption of adatoms at the edges of the MoS$_2$ nanoribbon gives rise to properties rather different from those in 1H-MoS$_2$.

\subsection{Vacancy Defects in MoS$_2$NR}

We investigated five different types of vacancy defects, namely Mo-, S-vacancy, MoS-, S$_2$-divacancy and MoS$_2$-triple vacancy, which are formed near the center of hydrogen passivated armchair ($n=12$) and zigzag ($n=6$) nanoribbons. All structures are optimized after the creation of each type of vacancy. Vacancy energies $E_V$, are calculated by subtracting the sum of the total energy of a structure having a particular vacancy type and the total energy(ies) of missing atoms in the vacancy defect from the total energy of the perfect structure (without vacancy). Here all structures are optimized in their ground states (whether magnetic or nonmagnetic). Positive $E_V$ indicates that the formation of vacancy defect is an endothermic process. In ~\ref{tab:vacancy} calculated vacancy energies as defined above and their magnetic ground states are presented.

\begin{table}
\caption{Calculated vacancy energies $E_V$ (in eV),
magnetic moments $\mu$ (in $\mu_B$) of five different types of
vacancy defects, Mo, MoS, MoS$_2$, S, S$_2$ in A-MoS$_2$NR and Z-MoS$_2$NR. NM stands for nonmagnetic
state with net $\mu$=0. $E_{i}$, energies of localized states in the band gap. Localized states are measured from the top of the valence bands in electron volt. The occupied ones are indicated by bold numerals and their spin alignments are denoted by either $\uparrow$ or $\downarrow$. Nonmagnetic states have no spin alignments.}\label{tab:vacancy}
\begin{center}
\tiny
\begin{tabular}{ccccccc}
\hline  \hline
& Mo & MoS& MoS$_{2}$ & S & S$_{2}$\\
& $E_V$-$\mu$ & $E_V$-$\mu$ & $E_V$-$\mu$  & $E_V$-$\mu$ & $E_V$-$\mu$ \\
\hline

A-MoS$_2$NR& 16.92-NM & 17.47-NM & 22.94-2.00 & 5.82-NM  & 11.55-NM  & \\
$E_{i}$ & \textbf{0.09}, \textbf{0.11}, 0.35, 0.48 & \textbf{0.11}, 0.40, 0.49 & \textbf{0.02}$\uparrow$, \textbf{0.03}$\downarrow$, 0.33$\downarrow$, 0.34$\downarrow$, 0.50$\uparrow$ & - &- \\
\hline

Z-MoS$_2$NR&  15.78-8.06  & 16.41-8.66 & 22.02-8.67 & 5.09-8.61& 10.77-8.31 & \\
\hline  \hline
\end{tabular}
\end{center}
\end{table}

For a hydrogen saturated armchair nanoribbon (A-MoS$_2$NR), having width $n$=12, the vacancy defects are treated in a supercell geometry, where a single defect is repeated in every four unit cell. For this supercell configuration, vacancy-vacancy coupling becomes minute and the resulting defect states appear as flat bands. Similar to 1H-MoS$_2$, all the vacancy types have zero net magnetic moments, except MoS$_{2}$-triple vacancy, which has a net magnetic moment of $\mu=2 \mu_{B}$ per supercell. The nonmagnetic excited states associated with vacancy defects occur above $\sim$120 meV, and are derived from Mo-$4d$ and S-$3p$ orbitals around the vacancy.

Similar to the armchair nanoribbons, various vacancy and divacancy defects in the hydrogenated zigzag nanoribbon (Z-MoS$_2$NR), having width $n$=6 are treated in a supercell geometry, where a single defect is repeated in every eight unit cell. Calculated vacancy energies, net magnetic moments per cell are presented in ~\ref{tab:vacancy}. It is found that in the presence of a vacancy defect, such as MoS-divacancy, S$_{2}$-divacancy, Mo-vacancy and S-vacancy, the spin-polarization of the zigzag nanoribbons appears to be suppressed. For example, while defect free, zigzag nanoribbons are metallic and spin-polarized ground state with net magnetic moment of $\mu=2.24 \mu_B$ per double unit cell, the total magnetic moment of eight unit cell decreases to $\sim \mu=8.30 \mu_B$ per supercell from $\mu=8.96 \mu_B$ in the presence of vacancy defects. In particular, the net magnetic moment of MoS$_{2}$-triple vacancy appear to compensate for the edge magnetization of the zigzag MoS$_{2}$ NR to result in a net magnetic moment of 8.67 $\mu_B$ per supercell.

\section{Discussion and Conclusions}

The phonon dispersion of bare A-MoS$_2$NR with $n=12$ is calculated and stability of nanoribbons are ensured. Armchair nanoribbons are nonmagnetic direct band gap semiconductors; their energy band gaps vary with its width and termination of edge atoms with hydrogen, whereas zigzag nanoribbons are found to be ferromagnetic metals. The bare zigzag nanoribbon is found to be a half-metal. Both nanoribbons are stiff materials, but their in-plane stiffness are calculated to be less than half of those of graphene and BN.

The adsorption of adatoms and creation of vacancy defects in MoS$_2$ nanoribbons have crucial effects in the electronic and magnetic properties. We found that several adatoms can be adsorbed readily at diverse sites with significant binding energy. In this respect, MoS$_2$ appears to be a material, which is suitable for functionalization. Similarly, net magnetic moment can be achieved through the adsorption of Co adatoms to the nonmagnetic armchair nanoribbons. In addition to spin-polarization, significant charges are transferred to (or from) adatom.

While vacancy defects of S, S$_2$, Mo and MoS created in hydrogen passivated armchair nanoribbon do not induce any magnetic moment, the creation of MoS$_2$ triple vacancy results in a significant magnetic moment in the system. Vacancy creation in hydrogen passivated zigzag nanoribbons however suppresses the magnetic moment occurring at the edges of the nanoribbon and results in a decrease in the total magnetic moment of the system. Briefly, functionalization of MoS$_2$ nanoribbons through adatom adsorption and vacancy creation appears to be a promising way to extend the application of MoS$_2$.

\begin{acknowledgement}

This work is supported by TUBITAK through Grant No:104T537 and Grant No: 108T234. Part of the computational resources have been provided by UYBHM at Istanbul Technical University. S.C. acknowledges TUBA for partial support. We thank the DEISA Consortium (www.deisa.eu), funded through the EU FP7 project RI-222919, for support within the DEISA Extreme Computing Initiative.

\end{acknowledgement}

\bibliography{final}

\begin{mcitethebibliography}{90}
\bibitem[Pau()]{Pauli}
It should be kept in mind that there are ambiguities in calculating charge
  transfer. In fact, different methods result in different values of charge
  transfer.\relax
\mciteBstWouldAddEndPunctfalse
\mciteSetBstMidEndSepPunct{\mcitedefaultmidpunct}
{}{\mcitedefaultseppunct}\relax
\EndOfBibitem
\end{mcitethebibliography}

\end{document}